\documentclass[12pt,epsf]{article}
\usepackage{amsmath}
\usepackage{amssymb}
\usepackage{bm}
\usepackage[dvips]{graphicx}

\begin{document}
\setlength{\oddsidemargin}{0cm}
\setlength{\baselineskip}{7mm}

\newcommand{\N}{{\mathcal N}}
\newcommand{\F}{{\mathcal F}}
\newcommand{\Tr}{\text{Tr}}
\newcommand{\tr}{\text{tr}}
\newcommand{\Str}{\text{Str}}

\newcommand{\thetabar}{\bar\theta}
\newcommand{\gammadot}{\dot{\gamma}}
\newcommand{\alphadot}{\dot{\alpha}}

\renewcommand{\theequation}{\thesection.\arabic{equation}}

\begin{titlepage}  \renewcommand{\thefootnote}{\fnsymbol{footnote}}
$\mbox{ }$
\begin{flushright}
\begin{tabular}{l}
KUNS-1905\\
hep-th/0403259\\

\end{tabular}
\end{flushright}

~~\\
~~\\
~~\\

\vspace*{0cm}
    \begin{Large}
       \vspace{2cm}
       \begin{center}
         {Non-planar Diagrams and Non-commutative Superspace in Dijkgraaf-Vafa theory}
\\
       \end{center}
    \end{Large}

  \vspace{1cm}

\begin{center}
          Takeshi M{\sc orita}\footnote
           {
e-mail address : takeshi@gauge.scphys.kyoto-u.ac.jp}

           {\it Department of Physics, Kyoto University,
Kyoto 606-8502, Japan}\\
\end{center}

\vfill

\begin{abstract}
\noindent 
We consider the field theory on non-commutative superspace and non-commutative spacetime that arises on D-branes in Type II superstring theory with a constant self-dual graviphoton and NS-NS $B$ field background.
$\N=1$ supersymmetric field theories on this non-commutative space (such theories are called $\N=1/2$ supersymmetric theories.) can be reduced to supermatrix models as in hep-th/0303210 \cite{KKM}.  
We take an appropriate commutative limit in these theories and show that holomorphic quantities in commutative field theories are equivalent to reduced models, including non-planar diagrams to which the graviphoton contributes.
This is a new derivation of Dijkgraaf-Vafa theory including non-planar diagrams.
\end{abstract}

\vfill
\end{titlepage}
\vfil\eject

\section{Introduction}
\setcounter{equation}{0}
It is generally interesting and difficult to study the $1/\hat{N}^2$ correction in large-$\hat{N}$ reduced models \cite{EK}.
We consider the $1/\hat{N}^2$ correction in Dijkgraaf-Vafa theory \cite{DV1,DV,DVGLZ,CDSW}, in which low energy effective theory of D=4 ${\mathcal N}=1$ supersymmetric gauge theory is equivalent to associated matrix model, in order to illuminate this problem in this paper.

The proof of Dijkgraaf-Vafa theory in $\N=1~U(N)$ gauge theory coupled to one adjoint matter is given in \cite{CDSW}.
It was shown there that the Schwinger-Dyson equations (the Konishi anomaly equations) of the field theory are equivalent to those of the associated matrix model for all holomorphic quantities.
As a result, the field theory is equivalent to the associated matrix model as far as holomorphic quantities are concerned.
The origin of this equivalence is shown in \cite{KKM, KKM2}.
This is a new large-$\hat{N}$ reduction in non-commutative superspace
\footnote{Rigorously, fermionic coordinates are non-anticommutative. However, we call them `non-commutative superspace' for simplicity.
In some case, we also use the term `non-commutative superspace' for `non-commutative superspace and non-commutative spacetime'.}.
As in \cite{NCYM},  field theories on non-commutative space $[x^\mu,x^\nu]=-iC^{\mu\nu} $ can be mapped to matrix models.
In this procedure, If the original field theories have supersymmetry, 
these can be described by superfields on a superspace coordinates $(x^\mu, \theta,\bar{\theta})$ and the corresponding matrix models are functions of $(\theta,\bar{\theta})$.
Then we can consider non-commutative superspace,
\begin{align}
\{\theta^\alpha, \theta^{\beta} \} = \gamma^{\alpha\beta},~~
\{\bar{\theta}^{\dot{\alpha}}, \bar{\theta}^{\dot{\beta}} \} = \bar{\gamma}^{\dot{\alpha}\dot{\beta}},
\label{NCSS01}
\end{align}
in the matrix models.
Therefore, we can map these matrix models to supermatrix models in which matrices are no longer functions of $(\theta,\bar{\theta})$.
We can derive the equivalence of Dijkgraaf-Vafa theory from these supermatrix models.
When we take these non-commutative parameters to zero in these theories, the field theories and supermatrix models are still equivalent when limited to the holomorphic quantities. 

In particular, in the holomorphic terms, the quantity $1/\hat{N}^2$ can be represented by the ratio of the non-commutative parameter of superspace $\gamma^{\alpha\beta}$ to that of spacetime $C^{\mu\nu}$ \cite{KKM}, 
\begin{align*}
\frac{g_m^2}{\hat{N}^2}=-\frac{64 \det \gamma }{(2\pi)^4 \det C},
\end{align*}
where, $g_m$ is an appropriate constant in the supermatrix model.
As a result, an expansion with respect to $1/\hat{N}^2$ in the supermatrix model can be naively regarded as that with respect to these non-commutative parameters in the non-commutative field theory.
However, construction of field theories on the non-commutative superspace (\ref{NCSS01}) is difficult and has not been achieved.
Therefore, the $1/\hat{N}^2$ expansion is meaningful only for the leading (planar) terms, which  
 correspond to the commutative field theory.
Thus, we can not consider non-planar quantities in the argument of \cite{KKM}.

In this paper, we will construct supermatrix models corresponding to field theories on non-commutative superspace,\cite{KKM, KKM2, TY, JP}
\begin{align}
&\{\theta^\alpha, \theta^{\beta} \} = \gamma^{\alpha\beta},~~
\{\bar{\theta}^{\dot{\alpha}}, \bar{\theta}^{\dot{\beta}} \} = 0, \nonumber \\
&[y^\mu, y^\nu] = -i C^{\mu\nu},
\label{NCSS02}
\end{align}
where $y^\mu=x^\mu + i \theta^\alpha \sigma^\mu_{\alpha\dot{\alpha}} \bar{\theta}^{\dot{\alpha}}$.
The construction of field theories on this non-commutative superspace has been achieved \cite{S} and these theories are called ${\mathcal N}=1/2$ supersymmetric theories.
Although this non-commutativity breaks the unitarity of the theory, we consider this theory on a Euclidean space and ignore this problem.
In section 2, we will show that amplitudes of non-planar diagrams disappear in usual supersymmetric field theories and appear in ${\mathcal N}=1/2$ supersymmetric theories.
When we take the commutative limit $C^{\mu\nu}\rightarrow 0$, $\gamma^{\alpha\beta}\rightarrow 0$, while holding the ratio $\det \gamma / \det C$ finite, the non-planar diagrams contribute to the commutative field theories.
In section 3, we will show that these higher genus quantities correspond to those of the supermatrix models.
Therefore, we will understand the equivalence between the commutative field theory and the supermatrix model including non-planar diagrams.
If we take the ratio $\det \gamma / \det C$ to 0, we obtain the usual commutative field theory to which the non-planar diagrams do not contribute. 

On the other hand, the non-commutative superspace \cite{S, OV,KPT,BGN} and non-commutative spacetime \cite{NCYM, GAO, SW} arises on D-branes in Type II superstring theory in constant self-dual graviphoton field strength $F^{\alpha\beta}$
 and constant NS-NS $B^{\mu\nu}$ background \cite{CCS}.
The non-commutative parameters are given by these background fields.
Then the quantity $1/\hat{N}^2$ in the reduced model is also expressed in terms of these background fields and the expansion with respect to $1/\hat{N}^2$ can be regarded as a development with respect to these fields.
Then, it is possible to take an appropriate commutative limit.
Under this limit, the commutative field theory exhibits finite  non-planar diagrams to which the graviphoton contributes.
This result reproduces analyses in \cite{OV,ACDGN}.

\section{Appearance of the non-planar diagrams in Dijkgraaf-Vafa theory}
\setcounter{equation}{0}

In this section, we calculate planar and non-planar diagrams in $\N=1~U(N)$ gauge theory coupled to one adjoint matter.
We show that the amplitudes of the non-planar diagrams disappear in commutative space \cite{DVGLZ} and do not disappear in non-commutative superspace \cite{TY}.
Especially, we will show that the amplitudes do not either disappear under the commutative limit $C^{\mu\nu} \rightarrow 0$, $\gamma^{\alpha\beta} \rightarrow 0$ with a fixed finite ratio $\det \gamma / \det C$.
We will interpret this result as the contributions of background graviphoton field strength and $B$ field.

\subsection{Diagram calculation in commutative superspace}
\label{21.1}
We will calculate planar and non-planar diagrams in $\N=1~U(N)$ theory.
The action is
\begin{align}
S= \int d^4 x d^2 \theta d^2 \bar{\theta}~ \Tr \left( e^{-V} \bar{\Phi} e^{V} \Phi \right)+ \int d^4 x d^2 \theta\left( \Tr~ W(\Phi) + 2\pi i \tau \Tr~ W^\alpha W_\alpha \right)+ c.c.,
\label{action3}
\end{align}
where $V$ denotes the vector superfield including the $U(N)$ gauge field,  $W_\alpha$ denotes its field strength,  $\Phi$ denotes a chiral superfield in the adjoint representation of $U(N)$ and $\tau$ denotes a gauge coupling constant.
$W(\Phi)$ denotes a $(m+1)$th order polynomial superpotential,
\begin{align}
W(\Phi)=\sum_{k=0}^m\frac{g_k}{k+1}\Phi^{k+1}.
\label{superpotential}
\end{align}
We can consider this potential in general, however it is enough to consider the simpler superpotential, 
\begin{align}
W(\Phi)=\frac{1}{2}m\Phi^2+\frac{1}{3}g \Phi^3,
\end{align}
in this section.

\begin{figure}
\begin{center}
\includegraphics[width=12cm, clip]{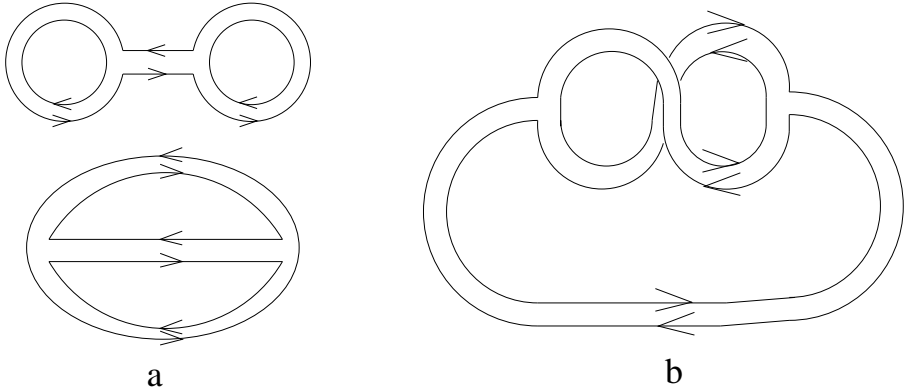}
\end{center}
\caption{(a) two-loop planar diagrams. (b) a two-loop $g=1$ non-planar diagram.}
\label{fig}
\end{figure}

First, we calculate two loop diagrams for matter field in this theory.
Since this theory is holomorphic, the matter kinetic term (D-term) and the superpotential (F-term) are decoupled.
Therefore, we can evaluate these amplitudes considering only the superpotential.
Then, the propagator of the superfield $\Phi$ is $1/m$ in terms of the holomorphic quantities.
The three point vertex is $g$.
Using these Feynman rules, we can calculate the two loop amplitude for matter field of figure 1 (a) as follows:
\begin{align}
(\frac{1}{2}+\frac{1}{6})\int \frac{d^4 k~ 4 d^2 \kappa}{(2\pi)^4}\frac{d^4 p~4 d^2 \pi}{(2\pi)^4}\left(\frac{1}{m}\right)^3g^2
=\frac{2g^2}{3m^3}\left(\delta^4(0)\delta^2(0)\right)^2.
\label{amplitude}
\end{align}
Here $\frac{1}{2}$ and $\frac{1}{6}$ are  symmetry factors, we omitted the traces and used usual and fermionic $\delta$ functions,
\begin{align}
\delta^4(x)=\int \frac{d^4k}{(2\pi)^4}e^{ikx},~~
\delta^2(\theta)= \int 4 d^2\kappa e^{-\theta \kappa}.
\end{align} 

A $\delta^4(0) \delta^2(0)$ singularity appears in equation (\ref{amplitude}) and we need to regularize it as follows \cite{KKM},
\begin{align}
\left.{\delta^i}_j\delta^4(y)\delta^2(\theta)\right|_{(y,\theta)\rightarrow(0,0)}= \frac{1}{64\pi^2}{\left(W^\alpha W_\alpha\right)^i}_j.
\label{K}
\end{align}
Here $i$ and $j$ are gauge indices.
This equation is a consequence of the Konishi anomaly \cite{Konishi}.
The gauge field contributes to the matter holomorphic terms only through this anomaly.
When we consider this theory as a low-energy theory of superstrings,
background graviphoton field strength and $B$ field do not contribute to this anomaly \cite{ACDGN}.

From the chiral ring properties \cite{CDSW},
\begin{align}
\{ W_\alpha, W_\beta \}=0,~~[\Phi,W_\alpha]=0,
\label{chiral}
\end{align}
amplitudes of the diagrams in which more than three $W^\alpha$ are in a single trace is zero.
Considering the combination of the three traces (there are three index loops) and four $W^\alpha$, we obtain
\begin{align}
\frac{2g^2}{3m^3}\left( 3N \frac{1}{64\pi^2} \Tr~ W^\alpha W_\alpha \frac{1}{64\pi^2} \Tr~ W^\beta W_\beta + 6\frac{1}{64\pi^2} \Tr~ W^\alpha W_\alpha \frac{1}{8\pi} \Tr~ W^\beta \frac{1}{8\pi} \Tr~ W_\beta \right).
\label{amplitude2}
\end{align}
Here $N$ is the rank of the gauge group and we simply assume that the gauge symmetry is not broken by the Higgs mechanism.

Next, we calculate a non-planar diagram (b).
The process is almost the same.
The difference is in the number of index loops.
This non-planar diagram has only one index loop. 
Since we must insert four $W_\alpha$ into one index loop,
this amplitude is zero because of the chiral ring properties.

 $\frac{1}{64\pi^2}\Tr~W^\alpha W_\alpha $ is replaced to glueball superfield $S$ in (\ref{amplitude2}) in the low energy theory.
Then we can obtain the two loop parts of the low energy effective action.
This result reproduces the calculus in \cite{DV1}\footnote{Dijkgraaf and Vafa calculate in $SU(N)$ theory and we do in $U(N)$ theory.}.
Therefore, Dijkgraaf-Vafa theory can be obtained from the calculation of the Feynman rules of the superfield $\Phi$,  Konishi anomaly (\ref{K}) and the chiral ring properties (\ref{chiral}).

\subsection{Diagram calculation in non-commutative superspace}
\label{21.2}

In this subsection, we calculate the two loop diagrams of figure 1 (a) and (b) in non-commutative superspace and non-commutative spacetime described by:
\begin{align}
\{\theta^\alpha, \theta^\beta \}&= \gamma^{\alpha\beta}, \nonumber \\
[y^\mu , y^\nu ]&= -iC^{\mu\nu}\nonumber, \\
\{ \bar{\theta}^{\dot{\alpha}}, \bar{\theta}^{\dot{\beta}} \}=\{ \theta^\alpha, \bar{\theta}^{\dot{\alpha}} \}=&[y^\mu, \theta^\alpha]=
[y^\mu, \bar{\theta}^{\dot{\alpha}}]=0.
\label{NCSS1}
\end{align}
Here $\gamma^{\alpha\beta}, C^{\mu\nu}$ are c-numbers.

The properties of the non-commutative superspace with $C^{\mu\nu}=0$ is studied in \cite{S}.
In the F-terms and the D-term of (\ref{action3}), 
we simply replace the standard products with star products \cite{AST} given by:
\begin{align}
f(y)*g(y)=\left. \exp \left(-\frac{i}{2}C^{\mu\nu}\frac{\partial}{\partial y^\mu} \frac{\partial}{\partial {y'}^\nu} \right)f(y)g(y') \right|_{y=y'},\label{star1} \\
f(\theta)\star g(\theta)=\left. \exp \left(-\frac{1}{2}\gamma^{\alpha\beta}\frac{\partial}{\partial \theta^\alpha} \frac{\partial}{\partial {\theta'}^\beta} \right)f(\theta)g(\theta') \right|_{\theta=\theta'}.
\label{star2} 
\end{align}
Although we need to treat the anti-holomorphic terms separately, 
it is not too serious a problem, since our interest lies in the holomorphic terms.

The holomorphy is broken on this non-commutative superspace \cite{S}.
The spacetime non-commutativity $C^{\mu\nu}$ does not prevent the holomorphy \cite{KKM}, but $\gamma^{\alpha\beta}$ does.
Therefore, when one take the commutative limit $\gamma^{\alpha\beta} \rightarrow 0$, $C^{\mu\nu} \rightarrow 0$ with the finite ratio $\det \gamma / \det C$, the holomorphy is recovered.
Since we are interested in field theories under the commutative limit, it is meaningful to consider the matter holomorphic terms in the non-commutative superspace as in the previous subsection.

Let us consider the Feynman rules of this non-commutative theory.
The propagator of $\Phi$ is the same: $1/m$.
As in the usual non-commutative field theory, the three point vertex exhibits a non-commutative phase \cite{TY},
\begin{align}
g e^{\left(-\frac{i}{2}C^{\mu\nu}k_\mu p_\nu -\frac{1}{2}\gamma^{\alpha \beta} \kappa_\alpha \pi_\beta \right)},  
\end{align}
where $k_\mu$ and $p_\mu$ are momenta and $\kappa_\alpha$ and $\pi_\beta$ are fermionic momenta. 
Since this non-commutative phase disappear in the planar diagrams,
the amplitude of the diagrams (a) is the same under the commutative limit $\gamma^{\alpha\beta}\rightarrow 0$, $C^{\mu\nu} \rightarrow 0$.
Note that one may regard the square of the $\delta$ function as $\det \gamma / \det C$ in (\ref{amplitude}) as we will show latter (\ref{square}) 
and one may derive another amplitude  proportional to $\det \gamma / \det C$.
However, this calculation is  non-physical, since this amplitude is $0$ when one takes $\det \gamma / \det C$ to $0$, and this result is inconsistent with the calculation of (\ref{amplitude2}).

Next, we consider the non-planar diagram (b).
In contrast to the planar diagrams, the non-commutative phase is not cancelled and the amplitude is
\begin{align}
\frac{1}{6}\Tr \int \frac{d^4 k~ 4 d^2 \kappa}{(2\pi)^4}\frac{d^4 p~4 d^2 \pi}{(2\pi)^4}\left(\frac{1}{m}\right)^3g^2 e^{\left(-iC^{\mu\nu}k_\mu p_\nu -\gamma^{\alpha\beta}\kappa_\alpha \pi_\beta  \right)}
&= \frac{g^2}{6m^3}\Tr\int \frac{d^4 k~ 4 d^2 \kappa}{(2\pi)^4} \delta^4(Ck)\delta^2(\gamma \kappa) \nonumber \\
&=\frac{g^2 N}{6m^3}\frac{4}{(2 \pi)^4} \frac{\det \gamma}{\det C}.
\end{align}
We can take the commutative limit while holding the ratio $\det \gamma / \det C$ finite and this amplitude does not vanish.

As a result, we obtain the amplitudes of the usual planar diagrams, as well as that of the non-planar diagram under the commutative limit of the non-commutative superspace.
What is the meaning of the non-vanishing amplitudes of the non-planar diagrams?
It is the remnant of the non-commutativity in the context of the field theory.
However, when we consider this theory as a low energy theory of superstring,
we can regard these amplitudes as the contributions of the background graviphoton field strength and NS-NS $B$ field.

The non-commutative superspace which we have considered arises on D-branes in Type II superstring theory in constant self-dual graviphoton field strength $F^{\alpha\beta}$ and constant NS-NS $B$ field background \cite{S,OV, BGN, SW} through, for example, calculation of hybrid formalism as in \cite{B}, where
\begin{align}
\{\theta^\alpha, \theta^\beta \}=&2{\alpha'}^2 F^{\alpha\beta} ,\label{NCSS2} \\
[y^\mu , y^\nu ]=&-i(2\pi\alpha')^2B^{\mu\nu},  \\
\{ \bar{\theta}^{\dot{\alpha}}, \bar{\theta}^{\dot{\beta}} \}=\{ \theta^\alpha, \bar{\theta}^{\dot{\alpha}} \}=&[y^\mu, \theta^\alpha]=
[y^\mu, \bar{\theta}^{\dot{\alpha}}]=0.
\end{align}
We take the limit $\alpha' \rightarrow 0$ while keeping a finite non-commutativity, 
so that the non-commutative parameters are related to $F^{\alpha\beta}$ and $B^{\mu\nu}$ as,
\begin{align*}
&\alpha' \longrightarrow 0,~~~~~~~ \\
&F^{\alpha\beta},~B^{\mu\nu} \longrightarrow \infty,\\
&(2\pi\alpha')^2B^{\mu\nu}=C^{\mu\nu}, \nonumber \\
&2{\alpha'}^2F^{\alpha\beta}=\gamma^{\alpha\beta}. 
\end{align*}
In this non-commutative superspace, we can derive a relation,
\begin{align}
\frac{ \det \gamma }{\det C}=\frac{4\det F}
{(2\pi)^{8} {\alpha'}^4 \det B}.
\label{F-N2}
\end{align}

Now, we can try to take the commutative limit.
If we simply take  $F^{\alpha\beta}$ and $B^{\mu\nu}$ to be finite,
the right side of (\ref{F-N2}) will diverge.
We need to take an appropriate limit to hold this ratio finite.
We choose to take the following limit,
\begin{align}
&B^{\mu\nu} \longrightarrow \sim (\alpha')^{-1}, \nonumber \\
&F^{\alpha\beta}~~~:\text{finite}, \nonumber \\
\Longrightarrow &~~~\gamma^{\alpha\beta} \rightarrow 0, C^{\mu\nu} \rightarrow 0 
\label{B-a}
\end{align}
then (\ref{F-N2}) is held finite as follows,
\begin{align}
\frac{ \det \gamma }{\det C} \sim \det F \sim F^2.
\label{F-N}
\end{align}
As a result, the factor of $\det \gamma / \det C$ in the amplitudes of the non-planar diagrams can be regarded as contributions of $F^2$ under this special limit.
The appearance of the amplitudes of non-planar diagrams proportional to the background self-dual graviphoton field strength $F^2$ has been argued by Ooguri and Vafa in \cite{OV}.
We will make some comments about relations with our study in section \ref{4}.

\section{The equivalence of the non-planar diagrams in Dijkgraaf-Vafa theory}
\label{2.1}
\setcounter{equation}{0}

We have shown that the amplitudes of the non-planar diagrams do not disappear.
In Dijkgraaf-Vafa theory \cite{KKM, DV1, DV, DVGLZ, CDSW, KKM2}, the amplitudes of the planar diagrams in the supersymmetric theory are equivalent to that of the corresponding matrix model.
We will show this equivalence can be maintained including the non-planar diagrams in general using the argument of Large-$\hat{N}$ reduction on the non-commutative superspace \cite{KKM, KKM2}.

\subsection{Field theory on non-commutative superspace and their reduced model}
We consider the action (\ref{action3}) with the general superpotential (\ref{superpotential}) on the non-commutative superspace (\ref{NCSS1}).
We map this field theory to a supermatrix model.
To do so, we introduce some matrices corresponding to the non-commutative superspace,
\begin{align}
&[ \hat{y}^\mu , \hat{y}^\nu] = -i C^{\mu\nu},
~~C^{\mu\lambda}B_{\lambda\nu}={\delta^\mu}_\nu,
~~\hat{p}_\mu = B_{\mu\nu}\hat{y}^\nu,
~~[ \hat{p}_\mu , \hat{p}_\nu] = iB_{\mu\nu},~~[\hat{y}^\mu,\hat{p}_\nu]=i \delta^\mu_\nu,
\label{NCcoordinate}
\\
&\{ \hat{\theta}^\alpha, \hat{\theta}^\beta \} = \gamma^{\alpha\beta},~~
\gamma^{\alpha\beta}\beta_{\beta\gamma}={\delta^\alpha}_{\gamma},~~
\hat{\pi}_\alpha=\hat{\theta}^{\beta}\beta_{\beta\alpha},~~
\{\hat{\pi}_\alpha,\hat{\pi}_\beta \}=\beta_{\alpha\beta},
~~ \{ \hat{\theta}^\alpha,\hat{\pi}_\beta \}=\delta^\alpha_\beta.
\end{align}
Then, fields on the non-commutative space correspond to matrices as follows \cite{KKM,NCYM, GAO},
\begin{align}
O(y)&=\int \frac{d^4 k}{(2\pi)^4} e^{ik_\mu y^\mu} \tilde{O}(k) &\leftrightarrow~~~~ \hat{O}
&=\int \frac{d^4 k}{(2\pi)^4} e^{ik_\mu \hat{y}^\mu} \tilde{O}(k), 
\label{Weyl2} \\
Q(\theta)&=\int 4d^2 \kappa~e^{-\theta^\alpha \kappa_\alpha}\tilde{Q}(\kappa) 
&\leftrightarrow~~~~ \hat{Q}
&=\int 4d^2\kappa~  e^{-\hat{\theta}^\alpha \kappa_\alpha}\tilde{Q}(\kappa) \nonumber \\
&= A + \theta^\alpha \psi_\alpha -(\theta^1\star\theta^2-\theta^2\star\theta^1) F  &
&= A + \hat{\theta}^\alpha \psi_\alpha - (\hat{\theta}^1\hat{\theta}^2-\hat{\theta}^2\hat{\theta}^1)F.
\label{Weyl}
\end{align}
The differential and integral operators are also mapped as follows, 
\begin{align}
&-i\partial_\mu O(y) \leftrightarrow [\hat{p}_\mu,\hat{O}],\\
&\int d^4y~\tr_{U(n)}O(y)=(2\pi)^2\sqrt{\det C}\Tr_{U(\hat{N})}( \hat{O} ),\label{int1}\\
&\frac{\partial}{\partial \theta^\alpha}O(y,\theta) \leftrightarrow [ \hat{\pi}_\alpha,\hat{O} \},\\
&\int d^2\theta~ Q(\theta) = \frac{i}{8\sqrt{\det \gamma }} \Str_{\theta} \left( \hat{Q} \right),\label{int2}
\end{align}
where $\Str$ denotes a supertrace defined as in \cite{KKM, KKM2}.
Then, we can reduce the action (\ref{action3}) to
\begin{align}
S=& \int d^2 \bar{\theta}~\frac{i(2\pi)^2\sqrt{\det C}}{8 \sqrt{\det \gamma}} \Str_{U(\hat{N})}\left( \hat{\bar{\Phi}}e^{\hat{V}}\hat{\Phi}e^{-\hat{V}}\right) \nonumber \\
&+ \frac{i(2\pi)^2\sqrt{\det C}}{8 \sqrt{\det \gamma}}\left\{ 2\pi i \tau \Str_{U(\hat{N})}(\hat{W}^\alpha \hat{W}_\alpha) + \Str_{U(\hat{N})}(W(\hat{\Phi}))\right\} \nonumber \\
&+ \int d^2 \bar{\theta}~(2\pi)^2\sqrt{\det C} \left\{- 2\pi i \bar{\tau} \Tr_{U(\hat{N})} \left( \hat{\bar{W}}_{\dot{\alpha}} \hat{\bar{W}}^{\dot{\alpha}} \right) +  \Tr_{U(\hat{N})} \left( \bar{W}(\hat{\bar{\Phi}}) \right) \right\}.
\label{action1}
\end{align}
Here, the hat indicates that the superfield is reduced as in (\ref{Weyl}) 
and their component fields are reduced as in (\ref{Weyl2}) 
\footnote{In the anti-holomorphic terms, we expand the anti-chiral superfields with respect to $\bar{y}$ and $\bar{\theta}$ and their component fields are mapped to matrices as in (\ref{Weyl2}) with respect to $\bar{y}$ instead of $y$. }.
The matter kinetic term and anti-holomorphic terms are functions of $\bar{\theta}$.
$\hat{N}$ is the infinite rank of the matrices and it is related to the bosonic non-commutativity $C^{\mu\nu}$ \cite{KKM}.
We introduce an appropriate dimensionful constant $g_m$ in the supermatrix model that is related to the non-commutative parameters through,
\begin{align}
\frac{\hat{N}}{g_m}=\frac{i(2\pi)^2 \sqrt{\det C}}{8\sqrt{\det \gamma}}.
\label{C-N}
\end{align}

We can construct in this way a reduced model of the gauge theory (\ref{action3}) in a non-commutative space (\ref{NCSS1}), which exhibits a different non-commutativity compared to the model (\ref{NCSS01}) which was studied in \cite{KKM}.

\subsection{Equivalence of the non-planar diagrams}
\label{2.2}
As in section 2,
the matter holomorphic terms of the action (\ref{action1}) are important to understand the holomorphic parts of the low energy effective theory.
Therefore, we discuss the action:
\begin{align}
S=\frac{\hat{N}}{g_m}\Str_{U(\hat{N})}(W(\hat{\Phi})),
\label{action4}
\end{align}
and consider the associated non-commutative field theory,
\begin{align}
\int d^4 x d^2 \theta~ \Tr~ W(\Phi).
\label{action7}
\end{align}

In this theory, we can show the equivalence of correlation functions:
\begin{align}
\left<\frac{g_m}{\hat{N}}\Str~\hat{\Phi}^k \right>
=
\left<\int d^4y d^2\theta~\Tr~\left( \delta^4(y)\delta^2(\theta)\delta^4(y)\delta^2(\theta)\Phi^k\right) \right>_{*\star}.
\label{map1}
\end{align}
Here we use (\ref{int1}), (\ref{int2}), (\ref{C-N}) and the equation \cite{KKM}:
\begin{align}
\delta^4(y)*\delta^4(y)\delta^2(\theta)\star\delta^2(\theta)=\frac{g_m^2}{\hat{N}^2}.
\label{square}
\end{align}
${*\star}$ on the right hand side of (\ref{map1}) indicates that we evaluate this amplitude in the non-commutative theory (\ref{action7}) as in section \ref{21.2}.
The left hand side is evaluated in the corresponding supermatrix model (\ref{action4}).
The left side can be expanded in powers of $g_m/\hat{N}$,
\begin{align}
&\left<\frac{g_m}{\hat{N}}\Str~\hat{\Phi}^k \right>=
\left<\frac{g_m}{\hat{N}}\Str~\hat{\Phi}^k \right>_0
+\left(\frac{g_m}{\hat{N}}\right)^2\left<\frac{g_m}{\hat{N}}\Str~\hat{\Phi}^k \right>_1
+\left(\frac{g_m}{\hat{N}}\right)^4\left<\frac{g_m}{\hat{N}}\Str~\hat{\Phi}^k \right>_2
+\cdots.
\end{align}
Here the lower right indices represent the contributions of the higher genus diagrams.
Correspondingly, this can be mapped to
\begin{align}
&\left<\int d^4y d^2\theta~\Tr~\left( \delta^4(y)\delta^2(\theta)\delta^4(y)\delta^2(\theta)\Phi^k\right) \right>_{*\star} \nonumber \\
&~~~~~~= \left<\int d^4y d^2\theta~\Tr~\left( \delta^4(y)\delta^2(\theta)\delta^4(y)\delta^2(\theta)\Phi^k\right) \right>_{{*\star}0} \nonumber \\
&~~~~~~+\left(\frac{8\sqrt{\det \gamma}}{i(2\pi)^2 \sqrt{\det C}}\right)^2
\left<\int d^4y d^2\theta~\Tr~\left( \delta^4(y)\delta^2(\theta)\delta^4(y)\delta^2(\theta)\Phi^k\right) \right>_{{*\star}1} \nonumber \\
&~~~~~~+\left(\frac{8\sqrt{\det \gamma}}{i(2\pi)^2 \sqrt{\det C}}\right)^4
\left<\int d^4y d^2\theta~\Tr~\left( \delta^4(y)\delta^2(\theta)\delta^4(y)\delta^2(\theta)\Phi^k\right) \right>_{{*\star}2}
+\cdots.
\end{align}
This means that the supermatrix model is equivalent to the non-commutative field theory for non-planar diagrams of genus $n$.
\begin{align}
\left<\frac{g_m}{\hat{N}}\Str~\hat{\Phi}^k \right>_n
=\left<\int d^4y d^2\theta~\Tr~\left( \delta^4(y)\delta^2(\theta)\delta^4(y)\delta^2(\theta)\Phi^k\right) \right>_{{*\star}n}.
\label{map3}
\end{align}

This equivalence has been established on the non-commutative superspace.
We are interested in studying this equivalence under the commutative limits,
$C^{\mu\nu} \rightarrow 0$,  
$\gamma^{\alpha\beta} \rightarrow 0$. 
When we take these limits,
a $\delta^4(0)\delta^2(0)$ singularity appears in equation (\ref{map3}) and we need to regularize it as in (\ref{K}).
Then we obtain
\begin{align}
\left<\frac{g_m}{\hat{N}}\Str~\hat{\Phi}^k \right>_n
=
\frac{1}{64\pi^2}\left<\Tr~ W^\alpha W_\alpha \Phi^k\right>_n .
\label{map2}
\end{align}
In this equation, 
the left hand side is of order $\left(g_m/\hat{N}\right)^{2n}$ and the right hand side is of order $\left(\det \gamma / \det C \right)^n$ compared to the leading order (planar diagrams).
Therefore, if $\det \gamma / \det C$ is finite, the contribution of non-planar diagrams in the field theory is finite, corresponding to the supermatrix model with finite $g_m/ \hat{N}$.

Using (\ref{F-N2}), (\ref{F-N}) and (\ref{C-N}), we can obtain,
\begin{align}
\frac{g_m^2}{\hat{N}^2}=-\frac{64 \det \gamma }{(2\pi)^4 \det C}=-\frac{(2)^8\det F}
{(2\pi)^{12} {\alpha'}^4 \det B} \sim \det F,
\end{align}
in the context of superstring background fields.
Therefore, we can regard the contributions of the non-planar diagrams in the supermatrix model as that of these fields.

This field theory is commutative but it is different from the usual commutative field theory\cite{DV1,DV} in which  the non-planar diagrams do not contribute to the amplitude. 
However, when we take the ratio $\det \gamma / \det C $ to zero, 
the contribution of non-planar diagrams disappear and our field theory can reproduce the calculation of the usual field theory.
In this sense, our theory can be regarded as an extension of Dijkgraaf-Vafa theory.

This result is consistent with the study of symmetries and mass dimension in \cite{CDSW, T}. 
In these papers, they calculate some charges and mass dimensions of operators and coupling constants, 
and they conclude that the symmetries forbid the non-planar diagrams to contribute to the holomorphic quantities in the supersymmetric gauge theory.
However, in our argument, we add the new constant $\det \gamma / \det C$ which also has these charges and mass dimension.
Therefore our calculation does not contradict their arguments.

Dijkgraaf-Vafa theory shows the equivalence of the prepotential $\F$ of this gauge theory and the free energy  $F_m$ of this matrix model\cite{DV1,DV}.
As in \cite{CDSW}, these quantities satisfy equations
\begin{align}
\frac{\partial \F|_{\psi=0}}{\partial g_k}&=\frac{1}{k+1}\frac{1}{64\pi^2}\left< \Tr~ W^\alpha W_\alpha \Phi^{k+1} \right>, \nonumber \\
\frac{\partial F_m}{\partial g_k}&=\frac{1}{k+1}\frac{g_m}{\hat{N}}\left< \Str~ \hat{\Phi}^{k+1} \right>,
\end{align}
where $g_k$ is a coupling constant in (\ref{superpotential}) and $\psi$ is a fermionic parameter in the prepotential of the $\N=1$ field theory\footnote{A relation between quantities calculated by supermatrix model and by bosonic matrix model is discussed in \cite{KKM} .}.
Since these equations hold including all diagrams, the equation (\ref{map2}) shows the equivalence of $\F$ and $F_m$ including non-planar diagrams.
As a result, Dijkgraaf-Vafa theory is applicable for non-planar diagrams as well.

\section{Conclusion and discussion}
\label{4}
\setcounter{equation}{0}

We have shown the equivalence between a field theory and a supermatrix model including non-planar diagrams and understood how the graviphoton field strength and $B$ field background contribute to these non-planar diagrams.
Our approach can be regarded as a new way to derive Dijkgraaf-Vafa theory from superstring theory.
It is interesting to compare our approach with the original Dijkgraaf-Vafa approach \cite{DV1,DV}.

Our argument is also applicable to field theories with gauge groups that are the products of some unitary groups coupled to adjoint, bifundamental and/or fundamental matter \cite{KKM2} and  we can study how the non-planar diagrams contribute to them. 

Our result that the graviphoton contributes to the non-planar diagrams can also be derived from arguments using diagrams \cite{OV} or Schwinger-Dyson equations \cite{ACDGN}.
These arguments use the C-deformation \cite{OV} :
\begin{align}
{\{W_\alpha, W_\beta \}^i}_j =  F_{\alpha\beta}{\delta^i}_j~~~\text{mod }~\bar{D},
\label{c-def}
\end{align}
and consider the theory on the commutative space.
(This deformation undoes the non-commutative superspace.)
The background fields in the superstring theory are different,
however our non-commutative superspace approach and this C-deformation approach give the same result in the field theory.
These two approaches should be related in some way. 

The meaning of the $1/\hat{N}^2$ correction is not clearly understood in general reduced models.
We have shown how the graviphoton and $B$ field, which are closed superstring background, contribute to the $1/\hat{N}^2$ corrections in our reduced model. 
It would be interesting to extend our approach to the graviton multiplet \cite{DGN} and propose some relation between closed string theory and reduced models.

The relation between non-planar diagrams and the graviphoton has also been advocated in the $\N=2$ field theory context \cite{GP}.
Our approach may be applied to these theories.

\section{acknowledgment}
We are grateful to T. Kuroki and M. Bagnoud for valuable advice and discussions on this paper.
We would also like to thank T.Azuma and H. Kawai for useful discussions.

T.M. is supported by a Grant-in-Aid for the 21st Century COE ``Center for Diversity and
Universality in Physics''.

\end{document}